# HONOLULU RAIL TRANSIT: INTERNATIONAL LESSONS FROM BARCELONA IN LINKING URBAN FORM, DESIGN, AND TRANSPORTATION

## Geoff Boeing
University of California – Berkeley, USA

The city of Honolulu, Hawaii is currently planning and developing a new rail transit system. While Honolulu has supportive density and topography for rail transit, questions remain about its ability to effectively integrate urban design and accessibility across the system. Every transit trip begins and ends with a walking trip from origins and to destinations: transportation planning must account for pedestrian safety, comfort, and access. Ildefons Cerdà's 19[th] century utopian plan for Barcelona's Eixample district produced a renowned, livable urban form. The Eixample, with its well-integrated rail transit, serves as a model of urban design, land use, transportation planning, and pedestrian-scaled streets working in synergy to produce accessibility. This study discusses the urban form of Honolulu and the history and planning of its new rail transit system. Then it reviews the history of Cerdà's plan for the Eixample and discusses its urban form and performance today. Finally it draws several lessons from Barcelona's urban design, accessibility, and rail transit planning and critically discusses their applicability to policy and design in Honolulu. This discussion is situated within wider debates around livable cities and social justice as it contributes several form and design lessons to the livability and accessibility literature while identifying potential concerns with privatization and displacement.

### Introduction

The Honolulu urban area is the fourth densest in the United States, trailing only those of Los Angeles, San Francisco-San Jose, and New York. Honolulu was the most traffic-congested U.S. city in 2011, ahead of stalwarts like Los Angeles and San Francisco (INRIX 2012). Yet unlike Los Angeles and San Francisco, Honolulu has not had an operational rail transit system to serve as an alternative to automobility since the early 20$^{th}$ century (Simpson and Brizdle 2000). It is, however, currently developing one. Honolulu's topography constrains the city to a long corridor along the coast, making rail development a sensible alternative. However, the new rail project may face challenges in integration and pedestrian access, given entrenched automobility and a sprawling urban periphery. Different researchers have theorized about investing in rail transit and its place within the complex transportation-land use connection. Cervero (1998) highlights three key preconditions for successful, sustainable rail transit systems: a strong city center, dense residential development, and long corridors of development in which to focus the rail lines. Downs (2004) also stresses the importance of density in rail catchment areas. Honolulu is presently a city dominated by the automobile (Kim 2010): despite its overall density, its periphery sprawls and its roads suffer from crippling congestion. Integration from institutional, operational, and physical perspectives is crucial for public transportation to compete with the automobile (Preston 2012). Integrating rail transit means luring drivers out of their cars (with carrots) and making drivers pay the full social cost of automobility (with sticks). It means strategic land use policy to shift origin-destination patterns and urban growth toward rail stations (Cervero 1998). Finally, it means supporting pedestrian access to transit, origins, and destinations through urban design and land use policy that fosters walkability (Lo et al. 2008). Every transit trip begins and ends with a walking trip *from* origins and *to* destinations: transportation planning must account for pedestrian safety, comfort, and accessibility.

This discussion is thus inherently situated within wider debates around livable cities. Livability has been theorized in innumerable ways since the dawn of urban planning. In the 19$^{th}$ century, Ildefons Cerdà – the father of modern Barcelona – developed a utopian theory of planning that emphasized the redistribution and homogenization of space to redistribute social differences and promote equality (Cerdà 1867). This theory of livability and the physical methods used to pursue it had enormous impacts on the urban form and human experience in Barcelona (Neuman 2011). Livability in the urban design literature today is commonly theorized as a bundle of interrelated characteristics linked to physical design that promote equity, stability, safety, comfort, walkability, accessibility, and community (Macdonald 2005; Bosselmann 2008). Livability is in turn nested within even broader debates around urban sustainability and justice, as it is inextricably dependent on the city's ability to meet all of its residents' ongoing needs into the future (Boeing et al. 2014). Several planning models – some competing, some complimentary – have taken

HONOLULU RAIL TRANSIT: INTERNATIONAL LESSONS FROM BARCELONA
IN LINKING URBAN FORM, DESIGN, AND TRANSPORTATION

up the mantle of livability in the U.S. today, including smart growth, the new urbanism, traditional neighborhood development, and transit-oriented development. Each promotes a compact urban form, walkability, and improved access to transit. Finally, issues of social justice cannot be ignored in the theorization of livability, as uneven distributions of power, capital, and privilege inevitably cloud the question of livability for whom and at the expense of whom (Evans 2002; Harvey 2010).

This article discusses policy and design lessons for Honolulu by examining the integration of rail into the livable urban form of Barcelona. It uses Barcelona's Eixample district – with its renowned walkability, well-integrated rail transit, and compact urban form – as a model of livability through design, density, land use, and pedestrian-scaled streets. Though their planning and political contexts differ in some ways, this study identifies several significant and transferable form and design lessons from the Eixample, including compact block sizes and height-to-width ratios that provide a sense of enclosure; streetscaping that balances visual complexity and order; mixed land uses; transit station assimilation into the urban fabric; and pedestrian equality within the circulation network. It also identifies potential concerns with affordability, privatization, and displacement, contributing to the livability and accessibility literature as well as to debates around social justice within it (Szibbo 2016).

In the following section, this article introduces the Honolulu context, including its physical and cultural settings, built environment, and transportation system. This includes a history of Hawaiian autocentrism and failed rail starts, as well as the current rail transit plans in Honolulu. Next it examines Barcelona's urban form – particularly Cerdà's top-down design of the Eixample district – and how it integrates rail transit with supportive pedestrian design and policy. Finally it discusses these findings – drawing several lessons from Barcelona's urban design, accessibility, and rail transit planning – and critically reflects on their applicability to policy and design in Honolulu.

## Honolulu and Rail Transit

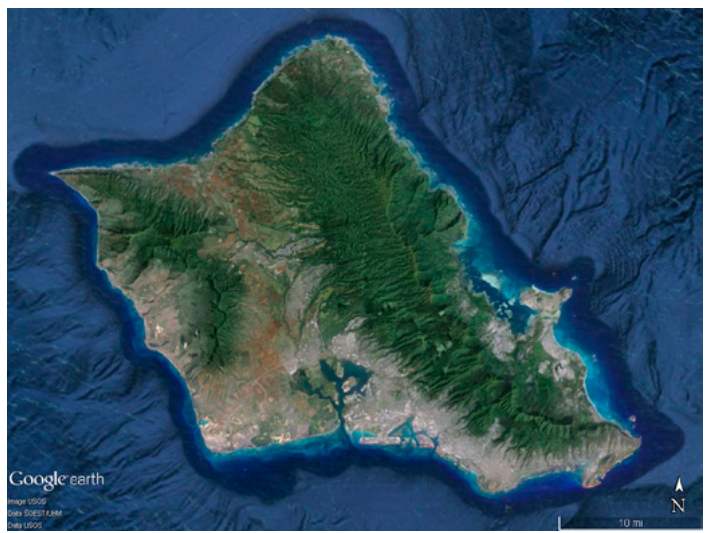

*Figure 1:*
The island of Oahu. The Honolulu metropolitan area is the urbanized region along the entire southern (bottom) portion of the island. Source: Google Earth.

Honolulu's transportation system is shaped by the region's natural, built, and sociodemographic characteristics. Honolulu is the largest city in Hawaii and the most remote major city in the world (Table 1). Located on the southern edge of the island of Oahu, metropolitan Honolulu is a long narrow strip of urban development pinned between the mountains and the sea. To the south is the Pacific Ocean. To the north lie the Koʻolau mountain range and the Waiʻanae mountain range, separated by the broad valley of the central Oahu plain between them, as shown in Figure 1. Urban development (and thus the transportation system) is constrained to a linear strip by the ocean and the mountains, although some development snakes its way up hillsides, through canyons, and along the central Oahu plain, as seen in Figure 2. Overall, the island of Oahu contains about 75 percent of Hawaii's population, 80 percent of which resides in the Honolulu metropolitan area.

Table 1: The most remote major cities in the world, by the great circle distance to the nearest city with at least 500,000 residents.

| Rank | City | Nearest Large City | Distance (km) |
|---|---|---|---|
| 1 | Honolulu, U.S. | San Francisco, U.S. | 3,854 |
| 2 | Auckland, New Zealand | Sydney, Australia | 2,156 |
| 3 | Perth, Australia | Adelaide, Australia | 2,131 |
| 4 | Anchorage, U.S. | Vancouver, Canada | 2,129 |

Due to its pleasant climate (National Weather Service 2014), diverse economy (Advameg 2009), attractive scenery, and relative wealth, Honolulu's population grew by an order of magnitude during the twentieth century (U.S. Census Bureau 2010). Due to its geographical constraints, the urban development to accommodate this influx has necessarily been dense by American standards: Honolulu is the fourth densest large urban area in the U.S. (Table 2) and has the nation's highest residential densities (Cervero and Duncan 2002). Further, it is a very diverse, white-minority city (U.S. Census Bureau 2010).

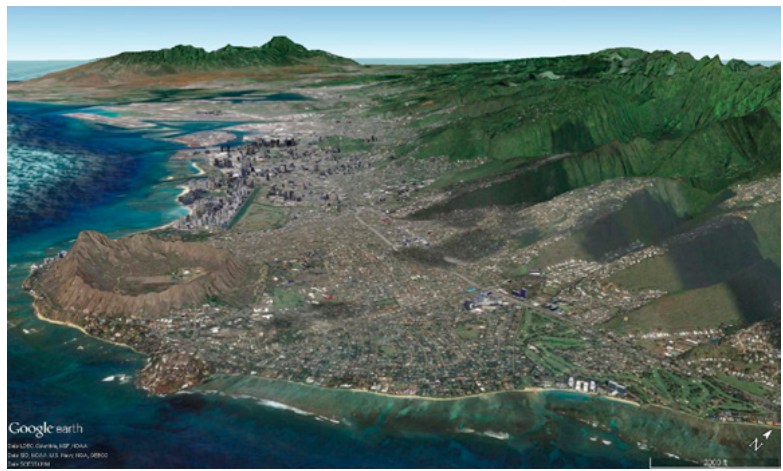

Figure 2: Honolulu is constrained by the mountains and the sea, and the city forms a linear corridor along Oahu's south coast. Source: Google Earth.

HONOLULU RAIL TRANSIT: INTERNATIONAL LESSONS FROM BARCELONA
IN LINKING URBAN FORM, DESIGN, AND TRANSPORTATION

The built environment in Honolulu is dense, spatially linear, and modern, with the fourth most high-rises of any U.S. city (Emporis 2014). It is also the most expensive metropolitan area in the country by regional price parity (U.S. Bureau of Economic Analysis 2014) and the second most expensive housing rental market (Gomes 2010). Race and class have come to the surface in recent gentrification debates as longtime residents and native Hawaiians are displaced from suddenly trendy districts by newcomers from the mainland and overseas. For instance, the formerly industrial Kakaako neighborhood – near downtown and along the proposed light rail alignment – was re-zoned residential/commercial in 1982 and has since seen a blitz of high-rise development and skyrocketing land values (Kavcic 2014). Issues of affordability, race, class, and social justice now indicate a challenge in critically engaging structures of power and capital in the pursuit of livability. Rising costs in newly thriving neighborhoods have pushed poorer, longtime residents – many of whom are native Hawaiian – toward the now-sprawling urban periphery in search of less expensive housing options.

Table 2:
The top four U.S. urban areas (among those with at least 100,000 people) by population density, according to the 2010 U.S. census. Source: U.S. Census Bureau

| Rank | Urban Area | 2010 Pop. | Land Area (sq km) | Density (per sq km) |
|---|---|---|---|---|
| 1 | Los Angeles | 12,150,996 | 4,496 | 2,703 |
| 2 | San Francisco-San Jose | 4,945,708 | 2,098 | 2,357 |
| 3 | New York | 18,351,295 | 8,935 | 2,054 |
| 4 | Honolulu | 802,459 | 440 | 1,824 |

Indeed, despite its overall density, Honolulu's current transportation system is quite auto-oriented like most other American metropolises. In 2000, Hawaii had approximately 744,000 cars, light trucks, and motorcycles compared to a civilian workforce of only 595,000 persons (Kim 2010). There is scant room available on its roads for all these vehicles, and as mentioned earlier, Honolulu ranked as the most traffic-congested city in the U.S. in 2011. The ensuing time costs, air pollution, pedestrian risks, and degraded urban design all harm livability for the sake of automobility. Honolulu's bus system – 'TheBus' – however has been a bright spot, providing a transit route within 0.8 kilometers of over 95% of the island's population. In 2010, it had the 6$^{th}$ highest ridership per capita in the U.S. and the lowest cost per passenger kilometer (Roman 2010). To complement this bus system, Honolulu officials recently began developing the first modern rail transit system in Hawaii.

*Honolulu High-Capacity Transit Corridor Project*
Honolulu is currently the densest urban area in the U.S. that lacks rail transit. Its traffic congestion, population density, topographical constraints, and mild climate have long made it an intriguing candidate for rail. Indeed, its politicians have tried – and failed until now – to build rail transit over the past 50 years. In 1966, Mayor Neil Blaisdell first proposed commuter rail as a solution to crippling workday traffic congestion. Mayor Frank Fasi initiated planning studies in 1977 for a rail project that eventually came to be known as Honolulu Area Rapid Transit and later the Honolulu Authority for Rapid Transportation (HART). However, President Ronald Reagan cut federal mass transit funding in the 1980s, leading to the project's cancellation in 1981. It was resumed in 1986, only to be permanently terminated by a 1992 city council vote against the tax increase needed to fund it (Epler 2014). Between 1994 and 2004, Mayor Jeremy Harris unsuccessfully pursued a bus rapid transit (BRT) project for the city (Levine 2012). Finally, Mufi Hannemann began the Honolulu High-Capacity Transit Corridor Project shortly after his 2004 election as mayor. Political battles raged in its wake (Genadio and Singh 2010). Detractors called for toll lanes or BRT instead, while proponents cited the democratic obligation to uphold a 2008 vote in favor of rail (Epler 2014). After several years of delays, ballots, petitions, voter referendums, and lawsuits, the rail project finally broke ground on February 22, 2011 (Park 2011).

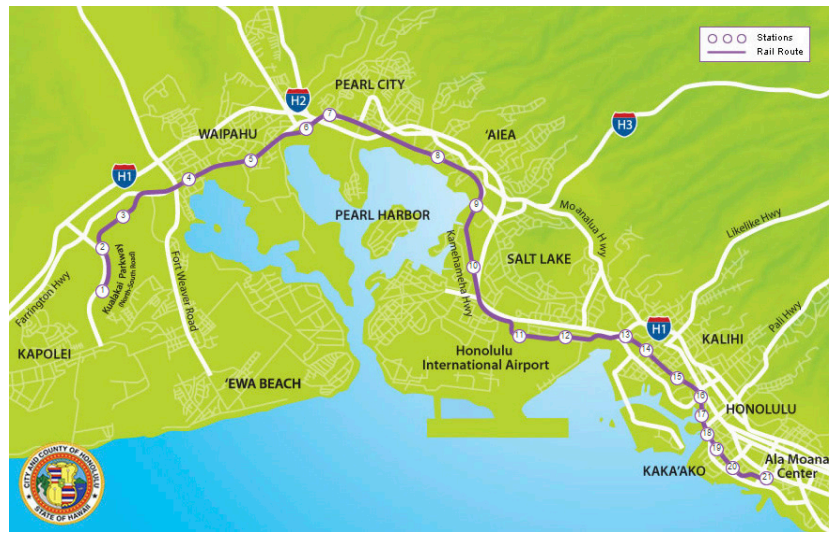

*Figure 3: Route map of the proposed Honolulu rail transit system. Source: Honolulu Rail Transit. Used with permission.*

The Honolulu Rail Project today is a 32-kilometer, $5.2 billion, elevated, driverless, rapid-transit light rail line connecting the western suburb of East Kapolei through Honolulu International Airport to downtown and on to the Ala Moana Center in the southeast. This route, shown in Figure 3, is expected to be completed by 2019. Future extensions are currently planned to also connect the University of Hawaii-Manoa and the posh, touristy Waikiki neighborhood at a later date (Genadio and Singh 2010; Epler 2014). Currently, about 73 percent of the total daily trips on Oahu originate between Kapolei and Waikiki. This corridor covers a significant portion of the city, alone containing over 60 percent of the entire island's population and 80 percent of its jobs. Nearly 100,000 people are forecast to ride the train daily, and individual trains will each hold over 300 people, departing every three minutes during rush hour (Epler 2014).

*Walkability and Pedestrian Access*

Every transit trip begins and ends with a walking trip *from* origins and *to* destinations. Honolulu's pedestrian and cyclist mode shares are currently suppressed by low peripheral housing costs inducing sprawl and a traditionally automobile-centric lifestyle in less urbanized areas. Additional and longer car trips are necessitated by Hawaii having the highest rate of private school attendance in the nation: at 16%, it is double the national average of 8%, and an astonishing 38% of students in Honolulu attend a private school (Wong 2014). Further, large swaths of Honolulu's current pedestrian street-level environment often poorly support livability and Hawaii consistently leads the nation in vehicular fatalities of elderly pedestrians (Wong 2012). The statewide pedestrian master plan highlights several areas of concern for pedestrians in Honolulu (State of Hawaii 2013a). These include sidewalk gaps, intersections without crosswalks, several unsignalized pedestrian crosswalks, intersections with high rates of pedestrian collisions, and districts with high concentrations of vulnerable populations, such as children and the elderly, who cannot drive.

At several intersections in Honolulu, the majority of pedestrian collisions occur while pedestrians are *within* the crosswalk. To improve walkability, the master plan includes a new pedestrian toolbox that recommends pedestrian and driver safety education programs, complete streets design, neighborhood design, maintenance, enforcement, and new physical guidelines for sidewalks, crosswalks, and intersections. Of particular interest, the toolbox addresses pedestrian access to transit through 'best practices for creating a seamless connection between pedestrian and transit modes of transportation' (ibid., p. 80). This covers the following topics: the importance of pedestrian access to transit; accessibility; transit in Hawaii; transit compatible planning and site design; coordination between agencies; transit-oriented development; transit stop locations; pedestrian routes to transit; intersections and crossings near transit; designing and improving transit facilities for good pedestrian access (State of Hawaii 2013b). Further, recent transit-oriented development proposals for Kakaako and Ala Moana – central neighborhoods just southeast of downtown – explicitly incorporate walkability, density, and complete streets into the planning of the light rail transit stations (RTKL Associates et al. 2014; Hawaii Community Development Authority 2015). Recent planning efforts focusing on infill, connectivity, and height and density bonuses are also promising.

These plans and toolbox present an excellent opportunity for Honolulu to improve its livability and pedestrian access to transit. Its current autocentrism poses dangers to pedestrians and promotes automobility at the expense of a more livable and balanced transportation mode split. Nevertheless, Honolulu faces several challenges in integrating its new rail system into the urban fabric. Every transit trip begins and ends with a walking trip and its current pedestrian environment is less than ideal. The American Institute of Architects (AIA) argues that the city's new elevated rail will only exacerbate the poor walking environment and harm livability by creating a cluttered and unattractive streetscape (AIA Honolulu 2011). Stakeholders have raised concerns about the aesthetics of the elevated trackway and the noise of passing trains (Camay et al. 2012). Supportive land use and urban design policy could be improved to better knit rail transit into the existing urban form and lifestyle – and to this end, we next consider how Barcelona integrates its transit system with a pleasant walking environment, supportive land use, and appealing urban design.

## Barcelona: Cerdà's Utopia?

Like Honolulu, Barcelona has a warm temperate climate and a large tourist industry (O'Sullivan 2014). It is similarly constrained by the Serra de Collserola mountain range on one side and the Mediterranean Sea on the other. Its urban area density of 2,356 persons per square kilometer (Institut d'Estadística de Catalunya 2014) would rank as the third densest in the U.S. – recall that Honolulu is currently fourth. Furthermore, Barcelona, like Honolulu, is a wealthy city with high per capita income. Barcelona is widely considered to be one of the world's most livable cities by urban designers, architects, and tourists. Its large Eixample district, designed in the 1850s according to the single-minded vision of planner Ildefons Cerdà, is particularly revelatory for our analysis. 'Barcelona today is unique; no other major European city owes its urban personality to the influence of a single individual… traffic flows better than in other cities of similar size, natural sunlighting, air circulation, and sanitation are better – ultimately, urban development is more rational' (Ordonez 1996, p. 20).

### Planning and Design of the Eixample

Cerdà designed the Eixample, meaning *extension* or *enlargement* in Catalan, in the 1850s when the medieval city walls around Barcelona's old town were torn down (Fernandez 2008; Casellas 2009). His utopian socialist planning ideology is encapsulated in his *General Theory of Urbanization* (Cerdà 1867). Cerdà theorized urbanism as a potentially equalizing force in a society suffering from inequality, and his key tool wielded toward this end was a rationalist and homogenous redistribution of space. He wanted this new city to be spacious, hygienic, and well-ventilated, with accessible green spaces woven throughout to improve urban livability particularly for the poor (Soria Y Puig 1995). The crowded, unsanitary old town had experienced several cholera epidemics in the 19[th] century and its population density (900 people per hectare) far exceeded those of London (100 people per hectare) or Paris and Madrid (300 people per hectare) (Serratosa 1996). Cerdà's plan spread out the city in a spatially homogenous manner intended to promote livability through social equality and uniformity.

HONOLULU RAIL TRANSIT: INTERNATIONAL LESSONS FROM BARCELONA
IN LINKING URBAN FORM, DESIGN, AND TRANSPORTATION

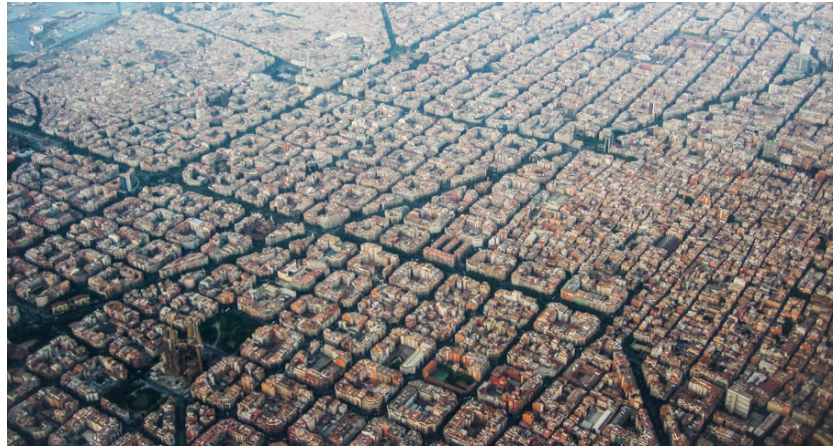

*Figure 4:*
The Eixample district from the air. "Eixample desde el avión" by alhzeia. Creative Commons BY-SA 2.0. Available at https://www.flickr.com/photos/ilak/. Accessed: 29/01/2016.

The Eixample is perhaps best characterized by its street pattern, seen in Figure 4. The grid's order and uniformity illustrate Cerdà's utopian theory of spatial redistribution to reduce inequalities (Illas 2012). While the old town comprises a dense, narrow, winding street network, the Eixample's streets are 20 meters wide and form unique octagonal blocks chamfered at each intersection. Each block is fairly compact at 113 x 113 meters and the grid is orthogonal and homogenous, punctuated only by a couple of major cross-cutting diagonal streets. Land use throughout the district is multifunctionalist and the building form is very consistent, characterized by perimeter blocks of approximately six-story buildings, about 20 meters in height, surrounding interior courtyards (Julia i Torne 1996). Though Cerdà sought a low-density urban form to reduce crowding, today's Eixample is considerably higher density due to ensuing and ongoing political and market pressures to build up (Cabre et al. 1996).

Cerdà was an early pioneer of both multimodal and intermodal transportation (Neuman 2009). His streets would explicitly accommodate casually strolling pedestrians, pedestrians carrying loads of goods, horse-drawn carriages, and a (never-realized) steam rail system (Magrinya 1996b). He assigned equal importance to traffic along the streets and to traffic entering and exiting buildings because he believed that 'mobility is only justified if it has a point of departure and a destination' (Serratosa 1996, p. 53). This early stance on the modern-day mobility versus accessibility debate (e.g., Levine et al. 2012) underscores the value Cerdà placed on pedestrian access in the Eixample. He also improved the pedestrian experience in the Eixample by concealing service networks such as water, sanitation, electricity, telegraph lines, and railways. In many cities, the street scene is visually cluttered with a mess of wires, poles, and towers. Cerdà deliberately concealed these service networks to create a more pleasant visual atmosphere for people spending time along his district's streets (Magrinya 1996a). The present-day undergrounded metro also circumvents the visual blight of overhead electricity wires or elevated tracks. Furthermore, the flexibility of Cerdà's scheme has been crucial as the Eixample has evolved in density and transportation technology, accommodating rail transit and automobiles over time as each subsequent mode was invented and then popularized (Perez et al. 2009; Guardia et al. 2010).

### The Eixample Today

The Eixample provides a valuable lesson in integrating transit with pedestrian access. Its relatively high density is conducive to ridership for rail transit and its thoughtful urban design and streetscaping make for pleasant walking trips between origins, public transit, and destinations. Specific supportive urban design features include wide sidewalks and pedestrian-friendly octagonal intersections. The streetscape includes mature street trees with broad canopies, ground floor retail, and consistent architectural quality with a high degree of visual complexity. Public furniture, public art, and elemental structures that protect residents from noise and pollution are ubiquitous parts of this highly livable pedestrian environment (Broto 2010).

Urban design scholar Allan Jacobs (1995) praises the livability of the Passeig de Gràcia, a major north-south boulevard running through the center of the Eixample in his book *Great Streets*. He specifically cites the 11-meter wide sidewalks and double row plantings of large plane trees, four to five stories tall, on either side of the roadway. The sidewalks are well-shaded by the massive, continuous tree canopy and they are paved with beautifully-detailed, Gaudí-designed tiles. These pedestrian areas are well-lit at night with attractive wrought-iron street lights. There are many Gaudí benches, positioned to face the public walkway. The buildings present a consistent yet diverse face to the public and feature many small stores with large windows to integrate them into the public realm. Shop entrances are typically less than 8 meters apart. In other words, the Passeig de Gràcia was well-designed for people to walk, and perhaps most germane to this discussion, it is well-integrated with rail transit: a metro line runs directly beneath this boulevard.

Metro construction in Barcelona began in the early 20[th] century (Busquets 2005). Today the city has three tram systems plus the metro system which consists of 163 stations and 11 lines that run underground in the central district but above ground in the suburbs. In the Eixample, transit stations are well-integrated into their block's context. They do not interrupt the urban form with parking lots, barren plazas, or elevated tracks but are rather tucked discretely right into the midst of numerous origin and destination points. Furthermore, the rail lines do not follow rights of way chosen only because they posed a path of least resistance to engineers. Rather, they traverse dense, high-amenity areas like the Passeig de Gràcia to knit the city together. Pedestrian access is also supported by a mixture of land uses across the district: residences, retail, professional offices, and cultural landmarks are distributed throughout the Eixample. Today, Barcelona has a transit mode share of 35 percent – very high for a wealthy Western city (Newman et al. 2009).

HONOLULU RAIL TRANSIT: INTERNATIONAL LESSONS FROM BARCELONA
IN LINKING URBAN FORM, DESIGN, AND TRANSPORTATION

Much of the Eixample's urban form, functionality, and livability today can be attributed to the seeds sown by Cerdà 150 years ago, but his original plans were neither wholly without flaw nor wholly successful. Although today's post-industrial Barcelona is noted for its multifunctionality, the Eixample originally developed during the industrial era and Cerdà (e.g. 1867) did not set aside land for industry. Instead, factories located at the outskirts, drawing the working-class toward the periphery and hindering his socially egalitarian motives. Ealham (2014) argues that Cerdà's utopianism manifested itself as bourgeois urbanism, entrusting too much to market forces and allowing landowners to undermine the original egalitarian goal of the Eixample. Despite its utopian socialist underpinnings, Cerdà's theory of urbanism sidestepped class conflict by focusing on the redistribution of space, while ignoring wealth and the contradictions of capitalism (Epps 2001, Harvey 2010). His plan, rather, hinged on persuading 'government officials and property owners to invest in urbanization while maintaining their status quo' (Illas 2012, p. 137). An egalitarian pursuit of livability was undermined by power and wealth, and the development of the Eixample succumbed to the pressures of politics and capital. Cerdà's blocks were originally planned to be lower density and bounded on only three sides, with public green space filling the center and open to the street. Over time, however, developers loosened these land use controls and doubled the heights of these blocks while sealing them with buildings on all sides, thus privatizing the now-enclosed central green spaces (Doerr 2014). Many of these green spaces were further converted to private parking lots in the 20[th] century, and the considerable width of certain streets in the district – originally intended to improve the flow of sunlight and air – sometimes feels too well-suited to the automobile today.

Yet all in all, the Eixample has fared well over time. Despite critiques, it has maintained considerable social diversity with a working and middle class while avoiding the gentrified banality of other successful cities. Barcelona's urban form has a strong, coherent identity and is highly legible and comprehensible (Bohigas 2004). The district features tree-lined streets with high-quality architecture and pedestrian-scaled visual complexity to make the whole trip from origin to station to destination enjoyable. The Eixample has fairly short blocks, its streets are 20 meters wide, and its buildings tend to be about 20 meters tall – creating an approximately 1:1 height-to-width ratio. Lastly, its service networks of wires, pipes, and rails are concealed to reduce unpleasant visual clutter.  Compactness, density, walkability, mass transit, and sustainability have long been integral to Barcelona's development (McDonough 2011). The Eixample integrates rail transit with pedestrian access by assimilating stations into the urban fabric and mixing land uses so that origins and destinations are near each other to foster walkability, and near rail stations to foster transit accessibility.

### Lessons in Linking Form, Design, and Transportation

Travel is a derived demand: outside of occasional joyrides and leisure strolls, people generally do not take a trip across town for the inherent sake of the trip itself. Rather, it is *access to things* that matters. Furthermore, every transit trip requires walking trips from the origin and to the destination. If these trips are unpleasant or poorly supported by urban design and policy, then transit ridership may diminish accordingly as travelers seek more agreeable routes and modes (Cervero 2013). While Honolulu's density, traffic congestion, topographical constraints, and climate make it a seemingly good candidate for rail transit, there are some points of concern.

First, Honolulu and Hawaii generally have a longstanding culture of automobility. Shifting modes to rail transit will require some degree of adaptation on the part of the residents. Cars provide the obvious benefits of privacy, speed (at least on uncongested roads), and flexibility in origin and destination. It can be challenging to lure drivers away from their vehicles, at a minimum because ingrained lifestyles and cultural connotations are difficult to change. Second, despite its overall density, Honolulu's built form has witnessed the same sprawling development patterns as the rest of the U.S. in the latter half of the 20$^{th}$ century. The benefits of its overall density will be diminished if walking trips to rail stations must be routed circuitously through a disconnected street network and faceless or even unpleasant urban design. Large, sprawling suburban areas such as Ewa, Mililani, and Hawaii Kai impose dependence on the private automobile and complicate regional public transportation patterns.

Third, the current rail plan is fairly expensive and may create an eyesore. Some critics have questioned the project's capital costs in light of its proposed ridership and capacity limits (O'Toole 2014). Its high costs could require high fares to recapture the investment, and high fares could hurt ridership. Its elevated route through the dense downtown could create visual blight, divide neighborhoods, and foster an unpleasant 'concrete jungle' pedestrian environment.  Fourth and finally, due to the first three points, walkability along parts of the new rail corridor could be less than ideal. All of these concerns have significant impacts for livability and maximum adoption of this enormous investment into rail transit. Honolulu's new pedestrian master plan and toolbox are good starting points to address these issues, but it is useful to compare other cities as well. While light rail in places such as Portland and Seattle have already been examined by Honolulu planners, it is also useful to consider an international context: Barcelona offers longstanding, invaluable lessons in integrating rail transit, pedestrian access, and livability. It serves as a representative case of the urban form and pedestrian experience that supports access to transit.

Reflecting on these two cities abstractly, any lessons must first acknowledge that the development of Honolulu and Barcelona exist in different sociopolitical contexts and

HONOLULU RAIL TRANSIT: INTERNATIONAL LESSONS FROM BARCELONA
IN LINKING URBAN FORM, DESIGN, AND TRANSPORTATION

planning eras. Cerdà's top-down socialist utopian theory and methodology have no corollary in Hawaiian planning practice today, and a new city cannot be cut from whole cloth. The Eixample was a 19th century utopian response to the social and physical ills of an overcrowded industrial era city. By contrast, Honolulu is a sprawling postindustrial city whose form is largely dictated by the automobile. There are considerable cultural differences between European urbanism and American suburbanism, including preferences for automobility. Honolulu's population is smaller, but like Barcelona, it anchors the major urban center in its region (Table 3). Its metropolitan density and wealth are fairly comparable to Barcelona's. Both cities have a similar warm but temperate climate with abundant sunshine. This type of climate both attracts tourists and fosters pleasant public spaces and walking trips – when the built environment cooperates to support them. Finally, both cities share a similar topography, constrained by mountains and the sea, which produces a fairly linear urban form conducive to rail transit and pedestrianism.

|  | Urban Area Population | Density (per sq km) | Metro GDP per capita | January High/Low Temperature (C) | August High/Low Temperature (C) |
|---|---|---|---|---|---|
| Honolulu | 802,459 | 1,824 | $59,968 | 26° / 18° | 31° / 24° |
| Barcelona | 4,702,008 | 2,356 | $34,821 | 14° / 5° | 29° / 21° |

Table 3: A comparative summary of Honolulu's and Barcelona's urban area population and density, metro GDP, and climate. Source: U.S. Census Bureau, Institut d'Estadística de Catalunya, World Meteorological Organization

To make its new rail system succeed, Honolulu will need to lure its residents out of their cars. In Hawaii as much as anywhere else, a coordinated and integrated approach is necessary. First, drivers should pay the full cost of their behavior. This means reducing the negative externalities from air pollution, congestion, and bundled and underpriced parking: only with a fair accounting of the full costs and benefits can mass transit compete against the automobile. Second, people switching to rail transit must enjoy their ride through the provision of comfortable, high quality train cars, stations, and service. Third, and the focus of this study, rail stations need to be accessible to pedestrians. Some rail systems in the U.S. attempt to balance the nature of rail stations as both a *node* in a transportation network as well as a *place* in and of themselves. As *nodes* in a larger intermodal network, stations may be surrounded by large park-and-ride lots to integrate with the predominately auto-focused transportation system. However, as *places* they need to be knit into the neighborhood, provide access to nearby amenities, and feature pleasant human-scaled design to attract pedestrians. Honolulu's density, climate, and traffic congestion would argue for more emphasis on this *place* paradigm to support livability – and recent local planning and TOD efforts seem to be moving in this direction.

Accordingly, Barcelona offers many urban form lessons as the Eixample integrates rail transit with pedestrian access in several ways. Its multifunctional mix of land uses ensures origins and destinations are near one another both for livability and for transit accessibility. A recent survey found that 79 percent of Americans believe they should walk more, but 40 percent do not do so because their neighborhoods lack nearby services, shops, schools, and workplaces (Kaiser Permanente 2013). Some of Honolulu's monolithic single-use residential sprawl could be improved by generating a greater diversity of uses and schedules, particularly around proposed rail stations. These stations should be well-integrated into the urban fabric, as they are in Barcelona, being neither an intrusive eyesore nor out-of-the-way and inconvenient. This observation lends credence to the AIA's argument that the elevated rail line through downtown Honolulu could create a visual disamenity. On one hand, elevated rail allows for rapid transit at a lower cost than an underground metro and at faster speeds than at-grade light rail. On the other hand, it creates something of an intrusive blemish that adds to the feeling of walking through a 'concrete jungle.'

Elevated rail is not necessarily ideal. This is especially true of the downtown, where underground rail might make more sense for short stretches (though Honolulu's soft alluvial soils and underground aquifers make it very expensive to build), or even at-grade segments that might be knit pleasantly into the surrounding streetscape (though this forfeits the speed benefits of grade separation). It is also unclear if light rail itself is necessarily the right choice. Barcelona underscores the importance of mass and density to generate a sufficient ridership base for rail. In an urban area of 802,000 persons, the flexibility and low cost of BRT might make even more sense to connect Honolulu's sprawling suburbs to its downtown. Already, the Honolulu bus system is planning for changes following new rail transit. According to TheBus's general manager Roger Morton, 'In areas the rail system will serve, what we will end up doing much more so than now is provide services to and from rail stations rather than provide long-haul services… we will provide a finer grain and a better level of service to all of the communities with more frequency, access and coverage' (Roman 2010).

Although it is less expensive than heavy rail or underground tunneling, light rail is still relatively expensive to build, especially when compared to BRT or otherwise expanded bus service. The first five U.S. cities to build light rail in the 1980s – Portland, San Diego, Sacramento, San Jose, and Buffalo – have had only mixed success. None of these cities today has a drastically higher transit mode share or a significantly larger center city share of total urban area population (Freemark 2014). However, each of these regions also built freeways in the past 30 years, largely failed to develop densely around transit, and did not charge local drivers the full cost of their automobility. As noted by Kim (2010) and Preston (2012), coordinated policy and integration are essential.

Regardless, rail development is already underway in Honolulu. To integrate it with pedestrian access, the city should look to the Eixample's high quality urban design. Its public furniture, interesting architecture, concealed service networks, and human-scaled visual complexity make walking from origins to stations to destinations pleasant. Shady tree-lined streets would also improve the hot and sunny pedestrian environment in Honolulu, and Cerdà's utopian plan progressively placed travelers of all types as equals in his streets. Autocentric Honolulu could improve livability and reduce pedestrian collisions by following similar principles of shared space and pedestrian equality.

The physical dimensions of the Eixample's urban form are also noteworthy. The blocks are not too long at 113 meters. The street widths and building heights are both 20 meters, creating an approximately 1:1 height-to-width ratio and a pleasant sense of enclosure for pedestrians. Cerdà's octagonal blocks create views and sightlines at intersections while also accommodating sunshine and airflow. Honolulu could improve the pedestrian experience in its denser districts by following this model. A city certainly cannot alter its urban form overnight, but incremental retrofitting and inevitable redevelopment can adhere to such design guidelines to improve the pedestrian experience and focus growth and density along the rail corridor. Finally, Honolulu can improve upon what Cerdà's vision failed to realize: first by protecting public space from enclosure and privatization, and second by acknowledging the roles of capital and power in urbanization and their repercussions on egalitarianism.

### Conclusion

This study explored the proposed rail transit system in Honolulu through the lens of Barcelona's Eixample district, and uncovered lessons for integrating rail transit with supportive pedestrian design and policy. It argued that the Eixample is a model of livability through design, density, land use, and pedestrian-scaled streets. Rail transit planning must account for pedestrian safety, comfort, and accessibility because every transit trip begins and ends with a walking trip. This study thus identified several significant form and design lessons from the Eixample, including compact block sizes and height-to-width ratios that provide a sense of enclosure; streetscaping that balances visual complexity and order; mixed land uses; transit station assimilation into the urban fabric; and pedestrian equality within the circulation network. Honolulu's climate and density make it a good city for walkability policies that support its rail investment, while topographical and natural resource constraints make it imperative to develop a sustainable transportation system integrated with a supportive urban form.

These findings contribute concrete examples of successful and supportive urban form in Barcelona to the livability and accessibility literature. They also identify potential concerns with privatization and displacement, arguing for a focus on social justice in the pursuit of livability. Sprawling, auto-dependent cities might benefit those families that can afford to live anywhere and purchase multiple vehicles, but it punishes everyone else through negative externalities, reduced accessibility, and degraded pedestrian environments. Incrementally embracing some of the Eixample's best urban form characteristics can democratize livability by encouraging transit adoption and improving the public realm. Cerdà's utopian theory and physical planning produced an urban form that is still considered highly livable today. However by ignoring crucial dynamics of power, privilege, and wealth, it may not be the success that it could have been. Today in Honolulu, gentrification around new rail investment sites threatens to create highly livable environments for a wealthy new urban elite at the expense of a displaced underclass. In many ways, today's smart growth and new urbanist paradigms represent a new urban utopianism that must critically engage with issues of wealth and power to become something more than mere spaces of displacement and gentrification. Thus this article argued that public planners must ensure that rail transit planning and investments in pedestrian-centric urban design produce livability for all citizens – not just a privileged few – by learning from Cerdà's successes as well as his mistakes.

This study focused on physical design and transportation planning, but future research could further explore questions of politics, praxis, and social justice. In the U.S., planners must be sensitive to how unwalkable many of its cities are. Americans have ingrained cultural preferences for automobility and sprawling residential communities: luring them out of their automobiles requires an attractive and efficient alternative mode of travel. Barcelona's lessons in density, urban design, and connecting rail with enjoyable pedestrian experiences are invaluable to Honolulu's livable and sustainable transportation future.

## References

› Advameg. (2009). Honolulu: Economy - Major Industries and Commercial Activity. Retrieved April 24, 2014, from http://www.city-data.com/us-cities/The-West/Honolulu-Economy.html
› AIA Honolulu. (2011). AIA Speaks Out on Transit. Retrieved April 24, 2014, from http://www.aiahonolulu.org/?261
› Boeing, G., Church, D., Hubbard, H., Mickens, J., & Rudis, L. (2014). LEED-ND and Livability Revisited. *Berkeley Planning Journal*, 27(1), 31–55.
› Bohigas, O. (2004). Ten Points for an Urban Methodology. In T. Marshall (Ed.), *Transforming Barcelona* (pp. 91–96). London: Routledge.

› Bosselmann, P. (2008). *Urban Transformation: Understanding City Design and Form.* Washington: Island Press.
› Broto, C. (2010). *City Embellishment: Barcelona*. Barcelona: Links International.
› Busquets, J. (2005). *Barcelona: The Urban Evolution of a Compact City*. Rovereto: Nicolodi.
› Cabre, A. M., & Munoz, F. M. (1996). Ildefons Cerdà and the Unbearable Density of Cities. In *Cerdà: Urbs i Territori: Planning Beyond the Urban* (pp. 37–46). Barcelona: Electa.
› Camay, S., Brown, L., & Makoid, M. (2012). Role of Social Media in Environmental Review Process of National Environmental Policy Act. *Transportation Research Record*, 2307, 99–107.
› Casellas, A. (2009). Barcelona's Urban Landscape: The Historical Making of a Tourist Product. *Journal of Urban History*, 35(6), 815–832.
› Cerdà, I. (1867). *Teoría General De La Urbanización* (2012 edition). Charleston: Nabu Press.
› Cervero, R. (1998). *The Transit Metropolis: A Global Inquiry*. Washington: Island Press.
› Cervero, R. B. (2013). Linking Urban Transport and Land Use in Developing Countries. *Journal of Transport and Land Use*, 6(1), 7–24.
› Cervero, R., & Duncan, M. (2002). Land value impacts of rail transit services in Los Angeles County. Report for the National Association of Realtors Urban Land Institute. Retrieved January 15, 2014, from https://drcog.org/documents/TODvalueLosangeles.pdf
› Doerr, A. (2014). Behind Four Walls: Barcelona's Lost Utopia. Retrieved January 15, 2014, from http://www.failedarchitecture.com/behind-four-walls-barcelonas-lost-utopia/
› Downs, A. (2004). *Still Stuck in Traffic: Coping with Peak-Hour Traffic Congestion*. Washington: Brookings Institution Press.
› Ealham, C. (2014). *Class, Culture and Conflict in Barcelona, 1898-1937*. London: Routledge.
› Emporis. (2014). Building data and construction projects worldwide. Retrieved April 24, 2014, from http://www.emporis.com/
› Epler, P. (2014). Honolulu Rail Project. Honolulu Civil Beat. Honolulu. Retrieved March 10, 2015, from http://www.civilbeat.com/topics/honolulu-rail-project/
› Epps, B. (2001). Modern Spaces: Building Barcelona. In J. R. Resina (Ed.), *Iberian Cities* (pp. 148–197). New York: Routledge.
› Evans, P. (2002). *Livable Cities: Urban Struggles for Livelihood and Sustainability*. Berkeley: University of California Press.
› Fernandez, V. A. (2008). Cerdà and Barcelona: Research and Plan. MIT, Cambridge, MA.
› Freemark, Y. (2014). Have U.S. Light Rail Systems Been Worth the Investment? Retrieved April 10, 2014, from http://www.citylab.com/commute/2014/04/have-us-light-rail-systems-been-worth-investment/8838/

- Genadio, F., & Singh, A. (2010). A Rough Ride on the Oʻahu Rail Transit Project. *Leadership and Management in Engineering*, 10(1), 21–31.
- Gomes, A. (2010). Honolulu Rents Still 2nd Priciest in U.S. *The Honolulu Advertiser*. Honolulu. Retrieved March 24, 2010 from http://the.honoluluadvertiser.com/
- Guardia, M., Garcia-Fuentes, J.-M., & Fava, N. (2010). The Construction of the Eixample in the Transformation of Contemporary Barcelona. In Urban Transformation: Controversies, Contrasts, and Challenges: Proceedings of the 14th IPHS Conference (pp. 473–487). Istanbul.
- Harvey, D. (2010). *Social Justice and the City*. Athens: University of Georgia Press.
- Hawaii Community Development Authority. (2015). Kakaako Community Transit Oriented Development Environmental Impact Statement. Honolulu. Retrieved March 20, 2015, from http://dbedt.hawaii.gov/hcda/kakaako-community-transit-oriented-development-draft-eis/
- Hawaii Department of Transportation. (2013a). Hawaii Pedestrian Toolbox. State of Hawaii: State of Hawaii Department of Transportation Highways Division.
- Hawaii Department of Transportation. (2013b). Statewide Pedestrian Master Plan. State of Hawaii: State of Hawaii Department of Transportation Highways Division.
- Illas, E. (2012). *Thinking Barcelona: Ideologies of a Global City*. Liverpool: Liverpool University Press.
- INRIX. (2012). National Traffic Scorecard Annual Report. Retrieved May 24, 2012, from http://www.inrix.com/scorecard/
- Institut d'Estadística de Catalunya. (2014). Idescat. Retrieved April 27, 2014, from http://www.idescat.cat/
- Jacobs, A. B. (1995). *Great Streets*. Cambridge: MIT Press.
- Julia i Torne, M. (1996). Barcelona's Municipal Building Ordinances and their Impact on the Eixample. In *Cerdà: Urbs i Territori: Planning Beyond the Urban* (pp. 61–66). Barcelona: Electa.
- Kaiser Permanente. (2013). Americans View Walking as Good for Health but Many Aren't Walking Enough to Realize Health Benefits. Retrieved March 10, 2015, from http://share.kaiserpermanente.org/article/americans-view-walking-as-good-for-health-but-many-arent-walking-enough-to-realize-health-benefits-survey/
- Kavcic, S. (2014). Kakaako Community for Seniors with Alzheimer's. University of Hawaii School of Architecture, Manoa, Honolulu, Hawaii.
- Kim, K. (2010). Transportation. In C. Howes & J. Osorio (Eds.), *The Value of Hawaii: Knowing the Past, Shaping the Future* (pp. 69–76). Honolulu: University of Hawaii Press.
- Levine, J., Grengs, J., Shen, Q., & Shen, Q. (2012). Does Accessibility Require Density or Speed? *Journal of the American Planning Association*, 78(2), 157–172.

HONOLULU RAIL TRANSIT: INTERNATIONAL LESSONS FROM BARCELONA
IN LINKING URBAN FORM, DESIGN, AND TRANSPORTATION

› Levine, M. (2012). Cayetano's Transit Plan Mirrors Harris' in 2000. *Honolulu Civil Beat*. Honolulu. Retrieved March 21, 2012, from http://www.civilbeat.com/articles/2012/03/21/15269-cayetano-transit-plan-mirrors-harris-in-2000/
› Lo, H. K., Tang, S., & Wang, D. Z. W. (2008). Managing the Accessibility on Mass Public Transit: The Case of Hong Kong. *Journal of Transport and Land Use*, 1(2), 23–49.
› Macdonald, E. (2005). Street-facing Dwelling Units and Livability: The Impacts of Emerging Building Types in Vancouver's New High-Density Residential Neighbourhoods. *Journal of Urban Design*, 10(1), 13–38.
› Magrinya, F. (1996a). Service Infrastructures in Cerdà's Urban Planning Proposals. In *Cerdà: Urbs i Territori: Planning Beyond the Urban* (pp. 189–204). Barcelona: Electa.
› Magrinya, F. (1996b). Way-Interways: A New Concept Proposed by Cerdà. In *Cerdà: Urbs i Territori: Planning Beyond the Urban* (pp. 205–224). Barcelona: Electa.
› McDonogh, G. W. (2011). Learning from Barcelona: Discourse, Power and Praxis in the Sustainable City. *City & Society*, 23(2), 135–153.
› National Weather Service. (2014). National Weather Service Forecast Office, Honolulu, HI. Retrieved April 24, 2014, from http://www.nws.noaa.gov/climate/xmacis.php?wfo=hnl
› Neuman, M. (2011). Ildefons Cerdà and the future of spatial planning: The network urbanism of a city planning pioneer. *Town Planning Review*, 82(2), 117–144.
› Newman, P., Beatley, T., & Boyer, H. (2009). *Resilient Cities: Responding to Peak Oil and Climate Change*. Washington: Island Press.
› Ordonez, J. A. F. (1996). Cerdà - Utopia and Pragmatism. In *Cerdà: Urbs i Territori: Planning Beyond the Urban* (pp. 19–22). Barcelona: Electa.
› O'Sullivan, F. (2014). Is Tourism Ruining Barcelona? Retrieved April 21, 2014, from http://www.citylab.com/work/2014/04/tourism-ruining-barcelona/8918/
› O'Toole, R. (2014). The Worst of Both: The Rise of High-Cost, Low-Capacity Rail Transit (Policy Analysis No. 750). Washington, D.C: Cato Institute.
› Park, G. (2011). It's a Go! *Honolulu Star-Advertiser*. Honolulu. Retrieved February 23, 2011, from http://www.staradvertiser.com/
› Perez, L., Medina-Ramon, M., Kunzli, N., Alastuey, A., Pey, J., Perez, N., … Sunyer, J. (2009). Size Fractionate Particulate Matter, Vehicle Traffic, and Case-Specific Daily Mortality in Barcelona, Spain. *Environmental Science Technology*, 43, 4707–4714.
› Preston, J. (2012). Integration for Seamless Transport. Presented at the Summit of the International Transport Forum in Leipzig, Germany: International Transport Forum Discussion Paper.
› Roman, A. (2010). Oahu Transit: Delivering Friendly Customer Service on the Island. *Metro*, 22–26.
› RTKL Associates, Belt Collins Hawaii, Fehr & Peers, & Keyser Marston Associates. (2014). Ala Moana Neighborhood Transit-Oriented Development Plan. Honolulu: City & County of Honolulu.

- Serratosa, A. (1996). The Value of Cerdà's Extension Today. In *Cerdà: Urbs i Territori: Planning Beyond the Urban* (pp. 47–60). Barcelona: Electa.
- Simpson, M., & Brizdle, J. (2000). *Streetcar Days in Honolulu*. Honolulu: JLB Press.
- Soria Y Puig, A. (1995). Ildefonso Cerdà's General Theory of "Urbanización." *Town Planning Review*, 66(1), 15–39.
- Szibbo, N. (2016). Lessons for LEED for Neighborhood Development, Social Equity, and Affordable Housing. *Journal of the American Planning Association*, 82(1), 37–49.
- U.S. Bureau of Economic Analysis. (2014). Real Personal Income for States and Metropolitan Areas, 2008-2012 (BEA 14-16) (pp. 1–9). Washington, D.C.: U.S. Department of Commerce.
- U.S. Census Bureau. (2010). American FactFinder. Retrieved April 23, 2014, from http://factfinder2.census.gov/faces/nav/jsf/pages/index.xhtml
- Wong, A. (2012). Don't Walk: Hawaii Pedestrians, Especially Elderly, Die at High Rate. *Honolulu Civil Beat*. Honolulu. Retrieved September 4, 2012, from http://www.civilbeat.com/articles/2012/09/04/17004-dont-walk-hawaii-pedestrians-especially-elderly-die-at-high-rate/
- Wong, A. (2014). Living Hawaii: Many Families Sacrifice to Put Kids in Private Schools. *Honolulu Civil Beat*. Honolulu. Retrieved March 17, 2014, from http://www.civilbeat.com/2014/03/21502-living-hawaii-many-families-sacrifice-to-put-kids-in-private-schools/